\documentclass[12pt, preprint]{aastex}  

\usepackage{graphicx} 
\usepackage{wrapfig} 
\usepackage{natbib}

\def\deg{$^{\circ}$~}
\def\arcsec{$\,^{\prime\prime}$~}
\def\arcmin{$\,^{\prime}$~}
\def\Msun{$M_\odot$~}
\newcommand{\Lx}{\ifmmode L_x \else $~L_x$\fi}
\newcommand{\lcgs}{\ifmmode erg~~s^{-1}\else erg~s$^{-1}$\fi}
\newcommand{\fcgs}{\ifmmode {\rm erg~cm}^{-2}~{\rm s}^{-1}\else
erg~cm$^{-2}$~s$^{-1}$\fi}
\newcommand{\mdot}{$\dot{{M}}$ }
 
\newcommand{\lsim }{{\lower0.8ex\hbox{$\buildrel <\over\sim$}}}
\newcommand{\gsim }{{\lower0.8ex\hbox{$\buildrel >\over\sim$}}}
\def\about{$\sim$}
\def\Lx{L$_x$~}

\begin{document} 

\title{\bf \Large Measuring the Accreting Stellar and 
Intermediate Mass Black Hole Populations in the Galaxy 
and Local Group} 

\title{White Paper Submitted to Astro2010 Science Frontier Panels\\
by\\
}

\normalsize

\author{Jonathan Grindlay (Harvard/CfA),  
Didier Barret (CESR-Toulouse),Tomaso Belloni (INAF-Brera), 
Stephane Corbel (Univ. Paris/Diderot), Phil Kaaret (Univ. Iowa), \\
Branden Allen (Harvard/CfA), Angela Bazzano (IASF-Rome), 
Edo Berger (Harvard/CfA), Govanni Bignami (IASF-Milan), 
Patrizia Caraveo (IASF-Milan),
Andrea De Luca (INAF-Milan), Pepi Fabbiano (SAO), Mark Finger
(NASA-MSFC), Marco Feroci (INAF-Rome), JaeSub Hong (Harvard/CfA), 
Garrett Jernigan (Berkeley), Michiel van der Klis (Univ. Amsterdam),  
Chryssa Kouveliotou (NASA-MSFC), 
Alexander Kutyrev (GSFC), Avi Loeb (Harvard/CfA),
Ada Paizis (INAF-Milan), Govanni Pareschi (INAF-Brera), 
Gerry Skinner (NASA-GSFC), Rosanne Di Stefano (SAO/CfA), 
Pietro Ubertini (IASF-Rome), and Colleen A. Wilson-Hodge
(NASA-MSFC) \\
}


\noindent{Contact person: Jonathan Grindlay, Department of Astronomy, 
Harvard University and Harvard-Smithsonian Center for Astrophysics 
University, \texttt{josh@head.cfa.harvard.edu}}

\vspace*{2.in}

\begin{center}
\noindent Submitted to {\underline {Science Frontier Panels:}}\\
{\bf Primary:} Stars and Stellar Evolution (SSE)\\
~~~~~~~~~~~~~~~~The Galactic Neighborhood (GAN) \\

{\bf Secondary:} Galaxies across Cosmic Time (GCT)
\end{center}

\newpage

\noindent{\bf Measuring the Accreting Stellar and 
Intermediate Mass Black Hole Populations in the Galaxy 
and Local Group} 

\vspace{5pt}
\noindent 
The population of stellar black holes (SBHs) in the Galaxy and 
galaxies generally is poorly known in both number and distribution. 
SBHs are the fossil record of the massive stars in 
galaxy evolution and may have produced some (if not all) of the 
intermediate mass (\gsim100\Msun) black holes (IMBHs) and, in turn, the 
central supermassive black holes (SMBHs) in galactic nuclei. 
For the first time, a Galaxy-wide 
census of accreting black holes, and their more readily recognizable tracer 
population, accreting neutron stars (NSs), can be measured. The key questions are: 
\begin{enumerate}
\item What are the distributions and numbers of accretion-powered 
SBHs (vs. NSs) in the Galaxy and IMBHs in the
Local Group?
\item Are ultra-luminous X-ray sources (ULXs) in nearby galaxies 
accreting IMBHs? 
\item How do accreting SBHs vs. NSs produce jets, and are isolated SBHs 
or IMBHs in giant molecular clouds detectable by their Bondi accretion?
\end{enumerate}


\noindent {\bf Tracing Stellar to Supermassive Black Holes}\\
It is increasingly clear that black holes are fundamental to the 
lives of galaxies and the energetics of the Universe. While the Big
Picture is dominated by the overwhelming role that supermassive black
holes (SMBHs) play in the formation and feedback on growth of central
bulges in galaxies, it is likely that the SMBHS are built in part 
by their stellar black hole (SBH) building blocks. The 
growth to SMBHs almost certainly includes mergers of SBHs in dense 
clusters (Portegies Zwart et al 2006) that may produce intermediate mass BHs 
(IMBHs), with masses $\sim100 - 1000$ \Msun. These can augment 
the mergers in galactic nuclei to build up SMBHs.  Likewise, the  
SBH population itself is partly driven by the 
neutron star (NS) population in galaxies and, particularly, 
in dense stellar clusters: NS-NS mergers to SBHs are then common, and 
produce short Gamma-ray bursts (Grindlay, Portegies Zwart and 
McMillan 2006)). 
The cascade of dependencies on lower mass compact objects as 
building blocks likely ends there, since accretion induced collapse 
to form NSs is less likely than SNIa's for accreting (or merging) 
white dwarfs. 

Tracing the hierarchical growth of SBHs to SMBHs 
can be initiated by the 
populations of SBH and IMBH systems that are themselves revealed 
when they accrete from binary companions. X-ray binaries, 
both with low mass and high mass donors (LMXBs and HMXBs, 
respectively), are then the markers 
for black hole seeds and evolution in galaxies. But the very 
accretion processes that make them visible render them very 
much fainter \gsim99\% of the time: accretion {\it instabilities} 
in their accretion disks shutoff the flow for either longterm 
deep quiescence or chaotic but usually short duty cycle transient 
outbursts. Thus the study of black hole demographics within galaxies 
demands the wide-field view and quasi-continuous monitoring 
that a hard X-ray imaging sky survey can provide as well as study. 

These same new, more sensitive, survey capabilities can also 
probe what could be the last reservoir of BHs and possibly IMBHs 
in the Galaxy: those which must be undergoing Bondi accretion 
from Giant Molecular Clouds (GMCs), without binary companions, 
to be revealed as hard X-ray sources.  With initial \lsim20\arcsec 
source positions and followup sensitive soft-medium X-ray  
and IR imaging (\lsim0.2\arcsec) and spectroscopy, 
background AGN and binary companion can 
be eliminated for the most promising candidates. These  
may include IMBHs postulated to be in the galactic halo as
remnant cores from sub-halo mergers (Volonteri and Perna 2005).

\vspace{5pt}
\noindent {\bf What we now know}\\
Over 20 accretion-powered SBHs 
in binary X-ray sources are confirmed in the Galaxy by 
\begin{wrapfigure}{r}{0.6\textwidth}
\centering \includegraphics[width=0.5\textwidth]{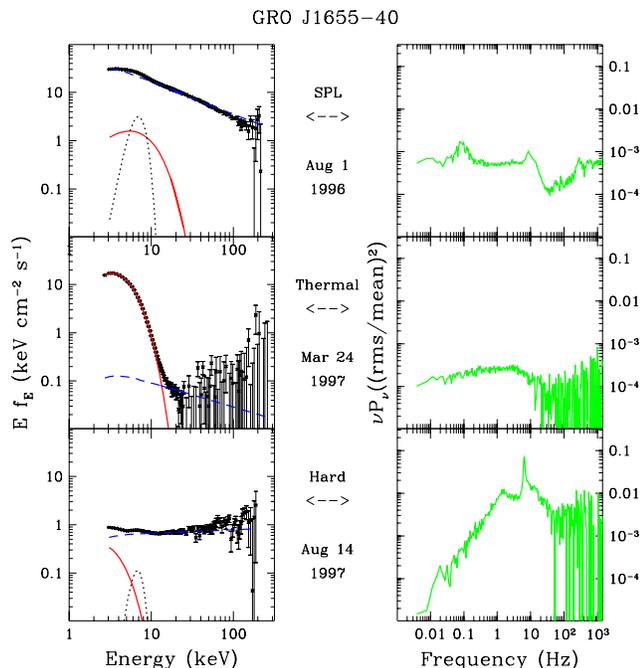}
\caption{{\it \footnotesize Spectral states (left) and corresponding 
power density spectra (right) for the three states 
(steep power law, thermal and hard) of the SBH GROJ1655-40 (from 
Remillard and McClintock 2006).}}
\noindent \hrulefill
\label{fig:mds}
\end{wrapfigure}
dynamical mass measurements of radial 
velocities of their secondary companions 
(Remillard and McClintock 2006). These SBHs have all been 
discovered as relatively luminous (\Lx \gsim10$^{36}$ \lcgs) 
and bright (\gsim100mCrab) sources. All but the original 
SBH prototype, Cyg X-1, 
are transients (though with differing duty cycles) and have 
been discovered at rates of $\sim0.5-1$/y ever since the bright 
prototype, A0620-00, was discovered (Elvis et al 1975). Most  
of these have had only 1-3 transient outbursts in the \about35 year 
history of moderately (in)complete sky coverage, so their typical 
recurrence time is probably \about20y. Another 20 sources, 
with similar X-ray properties (see below) and recurrence times 
are also probably SBHs though some (despite the absence of 
thermonuclear bursts, only possible on the solid surfaces of 
NSs) could be accreting NS-LMXBs. 

The galactic SBHs have relatively distinct X-ray 
spectral components, a soft thermal part associated with the 
accretion disk and one or more harder components which can 
extend to \gsim100 keV and whose origin is still 
controversial. Variations in these spectral components also  
correlate with short timescale variability properties and 
presence or absence of quasi periodic oscillations (QPOs). 
The low frequency 
QPOs, with frequencies generally 0.1 - 30Hz, correlate strongly 
with spectral state (Fig. 1) and can reach \about20\% rms 
power values, generally increasing with energy (up to the 
typically \about20 keV limits imposed by medium energy observations. 

These spectral-temporal properties are generally distinct 
from those for comparably luminous accreting NSs. The SBH systems 
are also strikingly correlated in their X-ray spectral 
\begin{wrapfigure}{r}{0.6\textwidth}
\centering \includegraphics[width=0.4\textwidth]{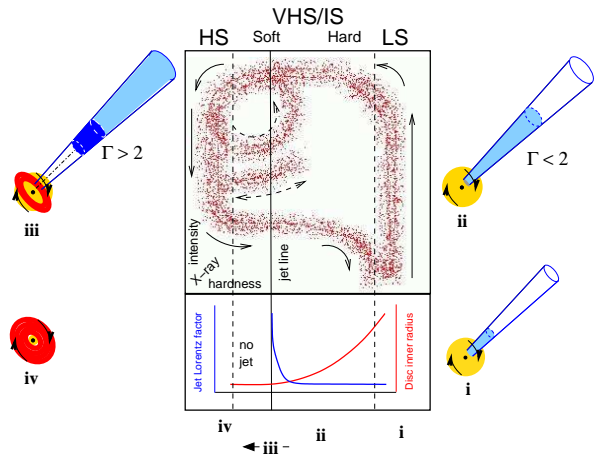}
\caption{{\it \footnotesize Hardness (ratio hard/soft flux) vs. 
intensity diagram, increasing upwards and to right, for observed 
behaviour of SBHs which produce radio and non-thermal jets when 
crossing the ``jet line'' (from Fender, Belloni and Gallo 2004).}}
\noindent \hrulefill
\label{fig:mds}
\end{wrapfigure} 
states with well defined transitions in the X-ray spectral hardness 
vs. intensity diagram (see Fig. 2), when crossing the ``jet line'',  
to production of non-thermal radio emitting jets, 
revealed by their power law spectra and spatially resolved emission. 
This striking jet transition is marked by timing and 
spectral changes which appear to scale with BH mass, and 
radio:X-ray coupling may in fact 
scale over 7 orders of magnitude to apply to AGN (Falcke, Kording 
and Markoff 2004). Compact jets also found in the low hard state 
with synchrotron spectra and turnover frequencies in the near-IR 
(Corbel and Fender  2002) are reminiscent of the self-absorbed 
jets considered (Blandford and Konigl 1979) for flat-spectrum AGNs.

Our knowledge of the spatial distribution of SBHs is limited by 
low statistics but nevertheless points to a disk and Bulge 
distribution. The long-recognized excess of luminous LMXBs 
(both NS and SBHs) in an ``extended'' (\about5\deg) Bulge distribution 
likely is connected to the radial distribution of much fainter 
(\Lx \about10$^{31-33}$ \lcgs) sources discovered initially with 
deep Chandra pointings (Muno et al 2003) and now traced to extend to 
much larger (\gsim4\deg) radii (Hong et al 2009). Although these 
sources likely are dominated by the more numerous accreting magnetic white 
dwarfs (CVs), a significant fraction are probably quiescent NS and SBH-LMXBs. 
There are no ULX sources, defined here as \Lx \gsim10$^{40}$ \lcgs,  
in the Galaxy and thus candidate IMBHs, though SS433 and possibly 
Cyg X-3 are contenders if beaming corrections are included.

The largest source of confusion in the SBH population 
is the much larger population of NSs in LMXBs in their low luminosity
(quiescent) states. These qLMXBs, undergoing minimal accretion 
in accordance with long term instability cycles, show hard 
PL spectra similar to SBH-LMXBs in their Hard state. This non-thermal 
spectral form for NS-qLMXBs is particularly dominant for 
accreting millisecond pulsars, which can be recognized by 
their coherent pulsations when in outburst. Their QPO power 
spectra are also separable (showing strong twin peak QPOs) from SBH 
systems, so that temporal analysis is needed to isolate the 
NS from SBH population. This distinction is possible in  wide-field 
hard X-ray imaging survey data for moderately bright (\about10-100mCrab) 
sources and QPOs are readily detected in bright (\lsim300 mCrab) 
sources even during scanning observations.

\vspace{5pt}

\noindent {\bf What is needed to disentangle the NS, SBH and IMBH
Populations?}\\
Given the ``language'' of accreting SBHs (Figs. 1 and 2), and the 
background ``sea'' of faint sources, which must include untold numbers 
of quiescent SBH-LMXBs (and NS-LMXBs), a sensitive hard X-ray imaging,
spectroscopic and timing survey is needed to unravel the source 
populations. In order to recognize transients with a range 
of duty cycles and durations, wide-field (full sky) coverage is 
needed on \lsim3h timescales. An energy band extending to at least 
100 keV can allow a complete characterization of the hard 
component, and lower energy limit of \about5keV ensures that the 
presence of the soft thermal component (cf. Fig. 1) can be established 
and its parameters measured. 
The primary discovery channel, hard X-ray imaging, 
must have \lsim5\arcmin spatial resolution 
to disentangle \gsim30mCrab sources in the crowded central Galactic 
Bulge and it must locate sources (when bright) to \lsim5-20\arcsec 
to enable optical/IR (OIR) identifications. It should be able to respond 
to outbursts which may last only \about1-2days (e.g. the peculiar 
hard X-ray ``nova'' CI-Cam (Belloni et al 1999) 
and conduct prompt OIR imaging and 
spectroscopy to directly identify the systems. And it should have 
high time resolution to distinguish the unique signatures of 
SBHs vs. NSs. These include the characteristic power 
spectra and QPOs of SBHs (Fig. 1) and sensitivity to 
X-ray bursts, and/or millisecond vs. longer period pulsations that 
immediately identify NSs accreting in LMXB or HMXB systems. 
Finally, to be able to detect ULX sources with 
\Lx \about10$^{40}$ \lcgs in Local Group galaxies 
at \about3-4 Mpc distances requires sensitivities 
\about0.5 - 1mCrab in the 20-100 keV band. This is 
comparable to that needed to detect SBHs in low-hard states or 
undergoing faint transient outbursts at \Lx \about5 x
10$^{34}$\lcgs in the galactic center region.  

In order to identify, and further characterize SBHs and IMBHs 
(detected as ULXs), the survey is further required to conduct higher 
sensitivity (narrow-field, focusing) imaging and spectroscopy 
at 0.1 - 10 keV to measure their low energy  absorption column (NH) 
to guide simultaneous OIR imaging and spectroscopy for source 
identification in obscured or crowded fields. The final OIR 
imaging and spectroscopy must be conducted with \lsim0.2\arcsec 
resolution for secure identifications. 

And finally, to isolate NSs from the SBH sample, photon event timing is 
required with \lsim0.1msec absolute timing to detect either 
coherent pulsations or kHz QPOs. For sufficiently faint 
sources, NS vs. SBH distinction may only be possible with the 
followup pointings of a sensitive soft-medium energy X-ray telescope 
(and OIR telescope) that are needed in any case for source 
identification. 

\vspace{5pt}

\noindent {\bf Science from a SBH/IMBH survey that could EXIST}\\
With a dedicated hard X-ray imaging survey, significantly 
more sensitive and thus able  
to speed up detections of SBHs (and NSs, and all other 
sources...) as compared to current wide-field galactic 
plane surveys with Swift/BAT  or INTEGRAL/IBIS 
(Kuulkers et al 2007), an exciting range of new science on 
demographics (and physics) of black holes in the Galaxy and 
Local Group becomes possible. 
Here we itemize some of the 
results that would likely arise from a survey with the instrument 
complement and mission currently in final phases of an Astrophysics 
Strategic Mission Concept (ASMC) Study: the newly-designed 
(for ASMC) Energetic X-ray Imaging Survey Telescope, {\bf EXIST}, 
see Appendix. 

\noindent
{\it SBH Spectral States Survey:}  \\
Every 3h (2 orbits), all known SBHs and all possible candidates 
with F(5-100keV) \gsim4mCrab, 
\begin{wrapfigure}{r}{0.5\textwidth}
\centering \includegraphics[width=0.4\textwidth]{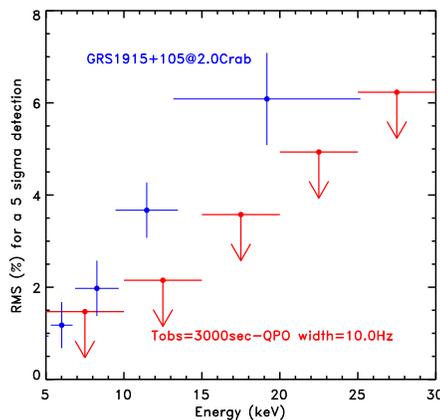}
\caption{{\it \footnotesize Sensitivity of EXIST (red) for a 3000s 
pointing or \about10 orbits of scanning coverage on the 
SBH GRS1915+105 and its 67Hz QPO, with \about10Hz width. 
The rms vs. energy actually observed with RXTE 
(Morgan et al 1997) are the blue points.}} 
\noindent \hrulefill
\label{fig:mds}
\end{wrapfigure}
corresponding to \Lx \gsim 3 X 10$^{35}$ \lcgs at the Galactic Center, 
would have  its spectrum measured and approximate spectral state (Figs. 1 and 2) 
established. Sources found to have very high states, with PL index
\about2.5, are likely SBH candidates, as may be  
those with steep/soft + hard components. 
Only Hard state (PLs with index \lsim2) are possible 
NS systems, which are identified either through the detection of 
pulsations/X-ray bursts or  by the combination of spectral and 
timing parameters. At low  
luminosities,  SBH and NS transients can be very similar and 
only distinguished if they show state transitions or X-ray  
bursts. {\it By continuous coverage with fine imaging 
and spectroscopy for 2y, and then nearly continuous (daily) 
coverage for 3y more, the full SBH catalog for the Galaxy will be 
assembled.}

\noindent
{\it SBH Temporal States Survey:} \\
The same survey observations just described will yield the first truly  
continuous sampling of timing properties of SBHs, which are intimately  
connected to the spectral variations. For the brightest sources, low- 
frequencies QPOs can be detected, but also the elusive high-frequency  
QPOs, probably directly connected to the Keplerian motion of matter  
close to the black-hole can be detected (as shown in Fig. 3 for GRS  
1915+105). NS QPOs are known to show a very different timing  
phenomenology and they will be easily identified in the sample.
The S/N for broader QPOs in SBHs can 
be improved by co-adding power spectra from several orbits or 
from the pointings that will occur for GRB and 
bright transient followups during the scanning survey mission 
phase (2y) and followup pointed mission phase (3y).

\noindent
{\it First continuous SBH vs. NS Survey:}\\
A key goal for the Galactic and Local Group BH survey is to detect 
and characterize transients as plausible SBH or IMBH candidates. 
For the Galaxy, the number of new SBH transients expected over 5y is 
at least 5-10 (conservatively), and more likely \gsim100. The lower value is 
straight extrapolation of the coverage of ``classical'' X-ray 
astronomy for the past 40y with luminous transients (\gsim1 Crab at 
peak; with typical decay times of ~30d) occurring at rates of 
\about0.5 - 1/yr. However EXIST would be at least 10X more 
sensitive and, equally important, has significantly larger 
duty cycle to be on source for short-timescale outbursts. EXIST 
would measure the (PL?) distribution of outburst amplitudes and 
durations of SBHs, which will constrain the total SBH population. 
In the past decade, accreting millisecond pulsars have been  
discovered and associated with relatively weak NS transients. The survey  
capabilities of EXIST will offer the unique combination of discovery  
of any outburst from new systems and the detection of pulsations by  
covering their full outburst.

\noindent
{\it First continuous ULX/IMBH Survey in Local Group:} \\
Both the 2y scanning and then 3y pointing mission phases of EXIST 
would allow the first continuous survey for outbursts of known IMBHs 
in the Local Group. The EXIST sensitivity is sufficient to detect the 
known ULX sources and candidate IMBHs, within 
M33 and M82, over 0.9mo and 2.8mo 
integration times for their quiescent flux values of \about0.8 and 
0.4mCrab (2-10 keV) corresponding to \Lx \about 9 x 10$^{39}$ \lcgs 
for both. The ULX  in Holmberg IX, with quiescent flux \about0.2mCrab 
and \Lx \about 3 x 10$^{39}$\lcgs could be detected  
on a 9mo timescale. Flaring outbursts from any of these will be 
detected on shorter timescales. These sources have only been ever 
observed sporadically with pointed soft X-ray telescopes (ROSAT,
Chandra and XMM) so little is known about their long term 
variability of flaring states.

\noindent
{\it First full-sky Survey for IMBHs and SBHs in GMCs:} \\
Finally, the long-discussed possibility (e.g. Grindlay 1978, 
Agol and Kamionkowski 2002) 
that isolated IMBHs could be detected undergoing Bondi accretion from  
giant molecular clouds (GMCs) in the galactic plane could be tested. 
For spherical accretion onto an IMBH of mass M moving at velocity 
V through a GMC with particle density n (cm$^{-3}$) of Hydrogen  
with proton mass m$_p$, the Bondi accretion rate is 
\mdot = 4$\pi$nm$_p$ (GM)$^2$/V$^3$. For a halo remnant IMBH 
with  velocity V = 100 km/s and mass M = 10$^3$ \Msun and 
GMC (core) gas density n = 10$^5$ cm$^{-3}$, this would 
then give an accretion luminosity, assuming a (conservative) 
accretion efficiency $\epsilon$ = 10$^{-4}$, 
\Lx \about5 x 10$^{35}$ $\epsilon_{-4}$ M$_{3}^2$/V$_{100}^3$, 
where subscripts denote the assumed scaling values. With a 2y full 
sky scanning survey sensitivity of 
Fx(5-100keV) \about0.06mCrab = 3 x 10$^{-12}$\fcgs, the corresponding 
distance such a IMBH-GMC could be detected is d \about40 kpc, so 
the entire Galaxy disk (and beyond) and GMC distribution could 
be surveyed. Given the likely small number of IMBHs in the 
Galaxy halo (\lsim10$^3$?) and small filling factor, no detections 
are likely. However for SBHs, with M = 10\Msun and V = 10km/s, 
the M$^2$/V$^3$ factor increases by a factor of 10 and so the survey 
for isolated SBHs in dense clouds becomes very feasible even for 
accretion efficiencies 1-2 orders of magnitude smaller. Thus the 
population of isolated stellar mass BHs passing through spiral arms and 
GMCs or dense clouds in the disk should be detectable as a population of hard 
sources with low energy cutoffs, and for which IR imaging and spectra 
would not reveal them to be LMXBs or background AGN.

\small
\noindent
{\bf Appendix: Brief Description of EXIST:}\\
EXIST (see http://exist.gsfc.nasa.gov/) is a LEO mission 
(600km, i =15/deg) with a primary wide-field survey 
High Energy Telescope (HET) employing a 4.5m$^2$ 
imaging (0.6mm pixels) Cd-Zn-Te (CZT) 
detector array sensitive over the 5-600 keV band. 
This  views the sky 
through a 7m$^2$ coded mask 2m above the detector plane to achieve 
2\arcmin imaging resolution and \lsim20\arcsec (90\% confidence radii) 
source positions within a 90\deg X 70\deg FoV. 
Two narrow-field imaging telescopes and
spectrometers are included to conduct the required prompt 
optical-IR and soft X-ray source identifications and followup 
studies. A 1.1m telescope optical-IR telescope (IRT),  
with simultaneous optical (0.3 - 0.9$\mu$m) 
and IR (0.9 - 2.3$\mu$) imaging (0.15\arcsec pixels) and spectroscopy 
(R = 30 or 3000), achieves high sensitivity (AB = 24 in 100s, enabled   
by cooling (-30C) the primary/secondary mirror 
to achieve zodiacal light-limited backgrounds. 
The Soft X-ray Imager (SXI), proposed to be contributed by 
Italy, is significantly more sensitive than the XRT on 
Swift. The proposed 5y mission would spend the first 2y continuously 
scanning the sky (full sky every 2 orbits, or 3h), interrupted by 
Gamma-ray Bursts (GRBs) \about2X per day which are imaged in 
real time (\about10-20sec) by HET to then slew the spacecraft to 
point the HET, SXI and IRT on the GRB position for source imaging 
identification and prompt redshift measurement. After a 2y 
scanning survey (HET and SXI), EXIST would be in pointed mode for the next 3y 
to followup with high sensitivity observations of survey sources.   
The HET would still cover \gsim90\% of the sky each day.
During the scanning mission, the full sky is imaged to 5$\sigma$ sensitivity 
F(5-100keV) \about4mCrab per 2 orbits, with total equivalent on-axis 
exposure \about8.4min on any source. For comparison with 
INTEGRAL/IBIS, which achieves F(20-60 keV) \about7 mCrab in 3.5h  
of on-axis exposure for a {\it single} 10\deg FoV 
(Kuulkers et al 2007), EXIST would reach the same sensivity 
{\it over the full-sky} every 8 orbits, or \about0.5day of elapsed (clock)
time.

\noindent
{\bf References}\\
Agol, E. and Kamionkowski, M. 2002, MNRAS, 334, 553\\
Belloni, T. et al 1999, ApJ, 527, 345\\
Blandford, R. and Konigl, A. 1979, ApJ, 232, 34\\
Corbel, S. and Fender, R. 2002, ApJ, 573, L35 \\
Elvis, M. et al 1975, Nature, 257, 656 \\
Falcke, H., Kording, E. and Markoff, S. 2004, A\&A, 414, 895\\
Fender, R., Belloni, T. and Gallo, E. 2004, MNRAS, 355, 1105\\
Grindlay, J. 1978, ApJ, 221, 234\\
Grindlay, J., Portegies Zwart, S. and McMillan, S. 2006, Nature Phys.,
2, 116 \\  
Hong, J., van den Berg, M., Grindlay, J. and Laycock, S. 2009, ApJ, submitted\\
Kuulkers, E. et al 2007, arXiv:astro-ph/0701244 and A\&A, in press \\
Morgan, E., Remillard, R. and Greiner, J. 1997, Ap.J., 482, 993 \\
Muno, M. et al 2003, ApJ, 589, 225\\
Portegies Zwart, S. et al 2006, ApJ, 641, 319\\
Remillard, R. and McClintock, J. 2006, ARA\&A, 44, 49 \\
Volonteri, M. and Perna, R. 2005, MNRAS, 358, 913


\end{document}